 \documentstyle[11pt,aaspp4]{article}

\newcommand{\Nco}{N_{\rm e}^*}		

\lefthead{Hirotani et al.}
\righthead{Pair Plasma Dominance in pc-scale Jet in 3C345}

\begin{document}

\title{PAIR PLASMA DOMINANCE IN THE PARSEC-SCALE RELATIVISTIC JET OF 3C345}

\author{Kouichi Hirotani, Satoru Iguchi,}
\affil{National Astronomical Observatory, Osawa, Mitaka 181-8588, Japan}

\author{Moritaka Kimura,\altaffilmark{1}}
\affil{Institute of Astronomy, The University of Tokyo, 
       Osawa, Mitaka 181-8588, Japan}

\and

\author{Kiyoaki Wajima}
\affil{The Institute of Space and Astronautical Science,
	Yoshino-dai, Sagamihara, Kanagawa 229-8510, Japan}

\altaffiltext{1}{present address: 
        National Astronomical Observatory, Osawa, Mitaka 181-8588, Japan}

\begin{abstract}
We investigate whether a pc-scale jet of 3C345
is dominated by a normal plasma or an electron-positron plasma.
We present a general condition that
a jet component becomes optically thick for synchrotron self-absorption,
by extending the method originally developed by Reynolds et al.
The general condition gives a lower limit of the electron number 
density, with the aid of the surface brightness condition,
which enables us to compute the magnetic field density.
Comparing the lower limit with another independent constraint 
for the electron density that is deduced from the kinetic luminosity,
we can distinguish the matter content.
We apply the procedure to the five components of 3C345 
(C2, C3, C4, C5, and C7)
of which angular diameters and radio fluxes at the peak frequencies 
were obtainable from literature.
Evaluating the representative values of Doppler beaming factors
by their equipartition values,
we find that all the five components are likely dominated
by an electron-positron plasma.
The conclusion does not depend on the lower cutoff energy of 
the power-law distribution of radiating particles.
\end{abstract}


\section{Introduction}

The study of extragalactic jets on parsec scales is 
astrophysically interesting in the context of 
the activities of the central engines of AGN.
In particlar, a determination of their matter content 
would be an important step in the study of jet formation, 
propagation and emission.
The two main candidates are a \lq normal plasma' consisting of
protons and relativistic electrons
(for numerical simulations of shock fronts in a VLBI jet, see
 G$\acute{\rm o}$mez et al. 1993, 1994a,b),
and a \lq pair plasma' consisting only of relativistic electrons
and positrons
(for theoretical studies of two-fluid concept, see
 Sol, Pelletier \& Ass$\acute{\rm e}$o 1989;
 Despringre \& Fraix-Burnet 1997).
Distinguishing between these possibilities is crucial for
understanding the physical processes occurring close to the 
central \lq engine' (presumably a supermassive black hole)
in the nucleus.

VLBI is uniquely suited to the study of the matter content of 
pc-scale jets,
because other observational techniques cannot image 
at milliarcsecond resolution and must resort to indirect means of
studying the active nucleus.
Recently, Reynolds et al. (1996) analyzed historical VLBI data
of M87 jet at 5 GHz (Pauliny-Toth et al. 1981)
and concluded that the core is probably dominated 
by an $e^\pm$ plasma.
In the analysis, they utilized the standard theory of synchrotron 
self-absorption to constrain the magnetic field, $B$ [G], and 
the proper number density of electrons, 
$\Nco$ [1/cm${}^3$] of the jet
and derived the following condition
for the core to be optically thick for self-absorption:
$ \Nco B^2 > 2 \delta_{\rm max}^{-2} $,
where $\delta_{\max}$ refers to the upper limit of the 
Doppler factor of the bulk motion.
This condition is, however, applicable only for the VLBI observations
of M87 core at epochs September 1972 and March 1973.
Therefore, in order to apply the analogous method to other AGN jets
or to M87 at other epochs, we must derive a more general condition.

On these grounds, Hirotani et al. (1999) generalized the condition
$ \Nco B^2 > 2 \delta_{\rm max}^{-2} $
and applied it to the 3C 279 jet on parsec scales. 
In that paper, they revealed that core and components C3 and C4,
of which spectra are obtained from the literature,
are dominated by a pair plasma.
It is interesting to note that the same concusion 
that 3C 279 jet is dominated by a pair plasma
is derived from an independent method by Wardle et al. (1998),
who studied the circularly polarized radio emission from 3C 279 jet.

In the present paper, we apply the same method to the 3C 345 jet.
The quasar 3C345 (redshift z=0.594) is one of a class of core-dominated 
flat-spectrum radio sources that are believed to emit 
X-rays via the synchrotron self-Compton (SSC) process.
VLBI imaging observations of the \lq\lq superluminal'' quasar 3C345
have been made at 5 GHz every year since 1977 
(Unwin \& Wehrle 1992)
while 10.5 and 22 GHz observations have occurred at more frequent
intervals (e.g., Biretta et al. 1986).
The apparent speeds of components C2, C3, C4, and C5
increase monotonically with time from $\sim 3c$ to $\sim 10 c$,
consistent with a jet of constant Lorentz factor ($\Gamma =10$)
bending away from the line of sight (Zensus, Cohen, \& Unwin 1995).
Later, Unwin et al. (1997) studied the time evolution of spectral shapes
and angular sizes of component C7 at a distance $\sim 0.5$ mas (2 pc)
from the nucleus.
Using the physical parameters given in the literature above,
and deducing the kinetic luminosity from its core-position offset,
we conclude that all the five jet components 
are likely dominated by an $e^\pm$ plasma.
In the final section, we discuss the validity of assumptions.

We use a Hubble constant $H_0 = 65 h$ km/s/Mpc and $q_0 =0.5$
throughout this paper.
These give a luminosity distance to 3C 345 of $D_{\rm L}=3.06h^{-1}$
Gpc.
An angular size or separation of $1$ mas corresponds to $5.83h^{-1}$ pc.
A proper motion of $1 \mbox{mas yr}^{-1}$ translates into a speed of 
$\beta_{\rm app}=30.3h^{-1}$.
Spectral index $\alpha$ is defined 
such that $S_\nu \propto \nu^\alpha$.

\section{Constraints on Magnetic Flux and Particle Number Densities}

We shall distinguish whether a radio-emitting component is 
dominated by a normal plasma or an $e^\pm$ plasma,
by imposing two independent constraints 
on $\Nco$.
First, in \S~2.1,
we give the synchrotron self-absorption constraint,
which is obtained by extending the work by Reynolds et al. (1996).
Secondly in \S~2.2,
the kinematic luminosity constraint is presented.

\subsection{Synchrotron Self-absorption Constraint}

In this paper, we model a jet component with redshift $z$ 
as homogeneous spheres of angular diameter $\theta_{\rm d}$, 
containing a tangled magnetic field $B$ [G]
and relativistic electrons which give a synchrotron spectrum with
optically thin index $\alpha$ and maximum flux density $S_{\rm m}$ [Jy]
at frequency $\nu_{\rm m}$.
We can then compute the magnetic field density as follows 
(Cohen 1985; Ghisellini et al. 1992):
\begin{equation}
  B = 10^{-5} b(\alpha) S_{\rm m}^{-2} 
      \left( \frac{\nu_{\rm m}}{\rm GHz} \right)^5 
      \left( \frac{\theta_{\rm d}}{\rm mas} \right)^4
      \frac{\delta}{1+z},
  \label{eq:lineA}
\end{equation}
where $\delta$ is the beaming factor defined by
\begin{equation}
  \delta \equiv \frac{1}{\Gamma(1-\beta\cos\varphi)} ,
  \label{eq:def-delta}
\end{equation}
$\Gamma \equiv 1 / \sqrt{1-\beta^2}$ is the bulk Lorentz factor
of the jet component moving with velocity $\beta c$,
and $\varphi$ is the orientation of the jet axis to the 
line of sight.
The coefficient $b(\alpha)$ is given in Cohen (1985).
Both $\Gamma$ and $\varphi$ can be uniquely computed from 
$\delta$ and $\beta_{\rm app}$ as follows:
\begin{equation}
  \Gamma = \frac{\beta_{\rm app}^2 +\delta^2 +1}{2\delta} ,
\end{equation}
\begin{equation}
  \varphi
  = \tan^{-1} \left( \frac{2\beta_{\rm app}}
		          {\beta_{\rm app}^2 +\delta^2 -1} \right) .
  \label{eq:solve_phi}
\end{equation}

We assume that the electron number density between
energies $E$ and $E+dE$ is expressed by a power law as
\begin{equation}
  \frac{dN_{\rm e}^*}{dE} = N_0 E^{2\alpha-1}.
  \label{eq:ele-dist}
\end{equation}
Integrating $d\Nco/dE$ from $\gamma_{\rm min} m_{\rm e}c^2$ 
to $\gamma_{\rm max} m_{\rm e}c^2$,
and assuming $\gamma_{\rm max} \gg \gamma_{\rm min}$
and $\alpha<0$,
we obtain the electron number density
\begin{equation}
  \Nco = \frac{\gamma_{\rm min}{}^{2\alpha}}{-2\alpha}
         (m_{\rm e}c^2)^{2\alpha} N_0.
  \label{eq:Ne-N0}
\end{equation}
Computing the optical depth along the line of sight,
Marscher (1983) expressed $N_0$ in terms of 
$\theta_{\rm d}$, $S_{\rm m}$, $\nu_{\rm m}$, and $\alpha$.
Combining the result with equation (\ref{eq:Ne-N0}),
we finally obtain (see also Appendix~B)
\begin{eqnarray}
   \Nco{}^{\rm (SSA)}
   &=&  e(\alpha) 
      \frac{\gamma_{\rm min}{}^{2\alpha}}{-2\alpha}
      \frac{h (1+z)^2 q_0{}^2 \sin\varphi}
           {zq_0 +(q_0-1)(-1+\sqrt{2q_0 z+1})}
   \nonumber \\
   &\times&
      \left( \frac{\theta_{\rm d}}{\rm mas} \right)^{4\alpha-7}
      \left( \frac{\nu_{\rm m}}{\rm GHz} \right)^{4\alpha-5} 
      S_{\rm m}{}^{-2\alpha+3}
      \left( \frac{\delta}{1+z} \right)^{2\alpha-3},
   \label{eq:Nmin}
\end{eqnarray}
where $e(\alpha) \equiv 2.39 \times 10^{1-6.77\alpha}$
($0<-\alpha<1.25$).
If the component is not resolved enough,
this equation gives the lower bound of $\Nco$.

\subsection{Kinetic luminosity constraint}

As described in Appendix~B in detail, 
we can infer the kinetic luminosity, $L_{\rm kin}$, 
from the core-position offset, $\Omega_{r\nu}$,
due to synchrotron self-absorption.
For the core, we assume a conical geometry with a small half opening 
angle $\chi$.
Then $L_{\rm kin}$ measured in the rest frame of the AGN becomes
\begin{eqnarray}
  L_{\rm kin}
  &\sim& C_{\rm kin} K 
      \frac{r_1{}^2}{r_\circ{}^3}
      \beta \Gamma (\Gamma-1) \chi^2
      \left( \frac{\Omega_{r\nu}/\nu_\circ}{r_1 \sin\varphi}
      \right)^{2(5-2\alpha)/(7-2\alpha)}
  \nonumber \\
  &\times&
      \left[ \pi C(\alpha) \frac{\chi}{\sin\varphi}
             \frac{K}{\gamma_{\rm min}}
             \frac{r_1}{r_\circ}
             \frac{-2\alpha}{\gamma_{\rm min}{}^{2\alpha}}
             \left(\frac{\delta}{1+z}\right)^{3/2-\alpha}
      \right]^{-4/(7-2\alpha)},
  \label{eq:Lkin_CPO}
\end{eqnarray}
where $K$ is defined by equation (\ref{eq:equiP_0})
and becomes $0.1$ for $\alpha=-0.5$ if an energy equipartition holds
between the radiating particles and the magnetic field.

For a pure pair plasma, we obtain
$C_{\rm kin}= \pi^2 \langle\gamma_-\rangle m_{\rm e}c^3/\gamma_{\rm min}$,
where $\langle\gamma_-\rangle$ is the averaged Lorentz factor
of randomly moving electrons and positrons,
which could be computed from equation (\ref{eq:ele-dist})
for a power-law distribution of radiating particles.

For a normal plasma, on the other hand, we obtain
$C_{\rm kin}= \pi^2 m_{\rm p}c^3/(2\gamma_{\rm min})$,
where $m_{\rm p}$ refers to the rest mass of a proton.
It should be noted that $\gamma_{\rm min}$ takes a different value
from a pair plasma.

Once $L_{\rm kin}$ of a stationary jet is obtained,
we can deduce $\Nco$ at an arbitrary position along the jet,
even if the geometry deviates from a cone.
When the jet has a perpendicular half width $R_\perp$ at a
certain position,
$L_{\rm kin}$ and $\Nco$ are related by
\begin{equation}
   L_{\rm kin}
   =  \pi R_\perp{}^2 \beta c \cdot \Gamma \Nco \cdot (\Gamma-1)
      \left( \langle\gamma_-\rangle m_{\rm e}c^2 
            +\langle\gamma_+\rangle m_+ c^2 
      \right) ,
   \label{eq:Lkin}
\end{equation}
where $\langle\gamma_-\rangle$ and 
$\langle\gamma_+\rangle$ refer to the 
averaged Lorentz factors of electrons and 
positively charged particles, respectively;
$m_+$ designates the mass of the positive charge.
Replacing the angular diameter distance,
$2R_\perp/\theta_{\rm d}$,
with the luminosity distance divided by $(1+z)^2$,
we can solve equation (\ref{eq:Lkin}) for $\Nco$ to obtain
\begin{eqnarray}
  \Nco{}^{\rm (kin)}
  &=& \frac{3.42 \, \times 10^2 h^2 q_0{}^4 (1+z)^4}
           {\left[ zq_0 +(q_0-1)(-1+\sqrt{2q_0 z+1}) \right]^2}
  \nonumber \\
  &\times& \left( \frac{\theta_{\rm d}}{\rm mas} \right)^{-2}
           \frac{1}{\beta\Gamma(\Gamma-1)}
           \frac{L_{46.5}}
                { \langle\gamma_-\rangle
                 +\langle\gamma_+\rangle m_+/m_{\rm e} }
           \mbox{\ cm}^{-3},
  \label{eq:Nco_Lkin}
\end{eqnarray} 
where $L_{46.5}$ refers to the kinetic luminosity
in the unit of $10^{46.5} \mbox{ergs s}^{-1}$.
It should be noted that 
$\langle\gamma_-\rangle +\langle\gamma_+\rangle m_+/m_{\rm e}$
becomes roughly 
$2 \gamma_{\rm min} \ln(\gamma_{\rm min}/\gamma_{\rm max})$
for a pair plasma with $\alpha \sim -0.5$,
while it becomes $1836$ for a normal plasma.
As a result, $\Nco{}^{\rm (kin)}$ for a pair plasma becomes about 
$100 \gamma_{\rm min}{}^{-1}$ times greater than that for a normal plasma.
Since $\Nco{}^{\rm (SSA)}$ is proportional to 
$\gamma_{\rm min}{}^{2\alpha}$,
the ratio $\Nco{}^{\rm (kin)}/\Nco{}^{\rm (SSA)}$ for a pair plasma
becomes about $100 \gamma_{\rm min}{}^{-1-2\alpha}$ 
times greater than that for a normal plasma.
For a jet component close to the VLBI core, 
we may put $\alpha \sim -0.5$;
therefore, the dependence on $\gamma_{\rm min}$ virtually vanishes.

In short, we can exclude the possibility of a normal plasma dominance
if $ 1 < \Nco{}^{\rm (pair)} / \Nco{}^{\rm (SSA)} \ll 100$
is satisfied,
where $\Nco{}^{\rm (pair)}$ refers to the value of 
$\Nco{}^{\rm (kin)}$ computed for a pair plasma.
On the other hands, $\Nco{}^{\rm (pair)} / \Nco{}^{\rm (SSA)} < 1$ 
implies that $L_{\rm kin}$ is underestimated.
The conclusion is invulnerable against the value of $\gamma_{\rm min}$
of electrons and positrons.

\section{Application to the 3C345 Jet}

Let us apply the method described above to the 3C345 jet on parsec
scales and investigate the matter content.
It is, however, difficult to define $\alpha$, $\nu_{\rm m}$, 
and $S_{\rm m}$ of each component well,
because the spectral information for an individual component is
limited by the frequency coverage and quality of VLBI measurements 
near a given epoch.
Therefore, Zensus et al. (1996) chose self-consistent values
that matched the data and gave a reasonable fit to the overall spectrum
when the components C2, C3, and C4 (hereafter, C2-C4), 
and the core are considered together (Table \ref{tbl-2}).
For C2 and C3 they used the highest value for $\nu_{\rm m}$, 
while for C4 they used a representative possibility.
Subsequently, Unwin et al. (1997) obtained these radio parameters
for C5 and C7 by analogous method.
We present these parameters together with their errors 
in Table \ref{tbl-3}.
The jet half opening angle $\chi \sim 2.4^\circ$ is calculated 
from measuring the jet size within $1$ mas distance
from the core (\S~4.3 in Lobanov 1998).
We choose $\alpha=-0.65$ as the spectral index of the core
below the turnover frequency at $700$ GHz
(\S 5.2 of Zensus et al. 1995).

\placetable{tbl-2}

\placetable{tbl-3}

\subsection{Kinetic Luminosity}

To estimate the kinetic luminosity from equation (\ref{eq:Lkin_CPO}), 
we have to input $\Gamma$, $\varphi$, $\Omega_{r\nu}$, and $\delta$
for a given $C_{\rm kin}$, $K$, $\chi$, and $\alpha$.
Let us first consider $\Gamma$, $\varphi$, and $\delta$.
As demonstrated in figure~4 in Unwin et al. (1995),
a component (C7) accelerated as it moved away from the core:
the Lorentz factor increased from $\Gamma \sim 5$ to $\Gamma > 10$,
and the viewing angle increased from $\varphi \sim 2^\circ$ to 
$\varphi \sim 10^\circ$.
It is inappropriate to consider the case $\varphi \ll \chi$; 
therefore, we assume $\varphi \sim 2^\circ$ for the core.
In this case, $\delta \gg 1$ holds to give
$L_{\rm kin} \propto \Gamma(\Gamma-1)/\delta \propto \delta$.
In the case of a newly born component (C7) at 1992.05,
Unwin et al. (1995) derived a conservative limit $\delta > 11.7$, 
by assuming that C7 was the origin of the observed X-rays.
Therefore, it is likely that $\delta$ is much greater 
than $10$ for the core, 
because $\delta$ decreased as the component moved away. 

The core-position offset of the 3C 345 jet 
was reported by Lobanov (1998),
who derived the reference value 
$\Omega_{r\nu}=10.7 \mbox{pc}\cdot\mbox{Hz}$.
For a pair plasma with $\alpha \sim -0.5$, 
$\langle \gamma_- \rangle \sim 
 \gamma_{\rm min} \ln(\gamma_{\rm max}/\gamma_{\rm min})$
holds in the expression of $C_{\rm kin}$;
therefore, equation (\ref{eq:Lkin_CPO}) gives 
\begin{equation}
  L_{\rm kin}
  \sim 10^{46} \frac{\ln(\gamma_{\rm max}/\gamma_{\rm min})}{10}
       K^{0.5}
       \left( \frac{\delta}{20} \right) \mbox{ergs s}^{-1}.
  \label{eq:Lkin_3C345_1}
\end{equation}
On the other hand, for a normal plasma, 
equation (\ref{eq:Lkin_CPO}) gives
\begin{equation}
  L_{\rm kin}
  \sim 10^{46} \left( \frac{\gamma_{\rm min}}{100} \right)^{-1}
       K^{0.5} 
       \left( \frac{\delta}{20} \right) \mbox{ergs s}^{-1}.
  \label{eq:Lkin_3C345_2}
\end{equation}
Unless the particles significantly dominates the magnetic field,
$K^{0.5}$ does not exceed unity
(see eqs. [\ref{eq:K_1}] and [\ref{eq:K_2}],
 which hold when an energy equipartition is realized between the
 radiating particles and the magnetic field).
For a normal plasma jet, the energy distribution must cut off
at $\gamma_{\rm min} \sim 100$ (\S~4; see also Wardle et al. 1998).
Since $\delta>100$ is unlikely for the 3C 345 jet,
we adopt $L_{\rm kin}=10^{46.5} \mbox{ergs s}^{-1}$
(or equivalently $L_{46.5}=1$) as the representative upper bound
in this paper.
If $L_{\rm kin}$ becomes less than this value, 
the possibility of normal plasma dominance further decreases.

\subsection{Equipartition Doppler factor}

We estimate the value of $\delta$ by assuming
an energy equipartition between the
magnetic field and the radiating particles.
In this case, $K$ becomes of the order of unity and 
$\delta$ is given by the so-called 
\lq\lq equipartition Doppler factor'' (Readhead 1994),
\begin{eqnarray}
  \delta 
  &=& \delta_{\rm eq} \nonumber \\
  &\equiv& \left\{ \left[ \frac{10^3F(\alpha)}
                               {(\theta_{\rm d}/{\rm mas})} 
	 	   \right]^{34}
	 	   \left[ \frac{2(h/1.54)}{1-1/\sqrt{1+z}} \right]^2
		   (1+z)^{15-2\alpha}
		   S_{\rm m}^{16}
                   \left( \frac{\nu_{\rm m}}{\rm MHz} 
                   \right)^{-35-2\alpha}
	   \right\}^{1/(13-2\alpha)} ,
  \label{eq:def-deleq}
\end{eqnarray}
where $F(\alpha)$ is given in Scott and Readhead (1977).

There is much justice in adopting the equipartition Doppler factor
as the representative value.
First, as G$\ddot{\rm u}$ijosa \& Daly (1996) pointed out,
$\delta_{\rm eq}$'s of various AGN jets have a high correlation with 
$\delta_{\rm min}$, 
the minimum allowed Doppler factor derived by comparing the predicted
and the observed self-Compton flux 
(Marscher 1983, 1987; Ghisellini et al. 1993).
(If a homogeneous moving sphere emits all the observed X-ray flux
via synchrotron self-Compton process, then $\delta$ equals 
$\delta_{\rm min}$.)
Secondly, the ratio $\delta_{\rm eq}/\delta$ depends weakly on 
the ratio $u_{\rm p}/u_{\rm B}$, 
where $u_{\rm p}$ and $u_{\rm B}$ refer to the energy densities of  
radiating particles (i.e., electrons and positrons) and the 
magnetic field, respectively.
For $\alpha=-0.75$ for instance, we obtain 
$\delta_{\rm eq}/\delta = (u_{\rm p}/u_{\rm B})^{2/17}$
(Readhead 1994).
It is noteworthy that $\Nco{}^{\rm (SSA)}$ depends relatively weakly on
$\theta_{\rm d}$, $\nu_{\rm m}$, and $\alpha$,
if we adopt $\delta=\delta_{\rm eq}$.
For example, we obtain 
$\Nco{}^{\rm (SSA)} \propto \theta_{\rm d}{}^{2.9} 
                        \nu_{\rm m}{}^{5.5}
                        S_{\rm m}{}^{-1.5}$
for $\alpha= -0.75$.
This forms a striking contrast with 
$\Nco{}^{\rm (SSA)} \propto \theta_{\rm d}{}^{-10} 
                        \nu_{\rm m}{}^{-8}
                        S_{\rm m}{}^{4.5}
                        \delta^{-4.5}$,
which would be obtained from equation (\ref{eq:Nmin})
without making any assumptions on $\delta$.

%
%

We present such representative values of $\delta_{\rm eq}$,
$\Gamma_{\rm eq} \equiv (\beta_{\rm app}^2 +\delta_{\rm eq}^2+1)
  /(2\delta_{\rm eq})$, $B$, and $\Nco{}^{\rm (SSA)}$ for C2-C4 
in Table \ref{tbl-2}, and those for C5 and C7 in Table \ref{tbl-3}.

We first compare the values of $\delta_{\rm eq}$ with 
$\delta_{\rm min}$.
It follows from Tables \ref{tbl-2} and \ref{tbl-3} that
$\delta_{\rm eq} > \delta_{\rm min}$ is satisfied for all the
eight cases, as expected.
Moreover, the values of $\delta_{\rm eq}$ 
for C2-C4 at 1982.0 and those for C7 at the four epochs,
decrease with increasing projected distance, $\rho$ [mas], from the core.
As a result, the viewing angle computed from $\beta_{\rm app}$
and $\delta_{\rm eq}$ (see eq. [\ref{eq:solve_phi}]),
$\varphi_{\rm eq}$,
increases with increasing $\rho$.
(We exclude C5, for which the trajectory appears in a different 
position angle from those for C2-C4.)
The results are qualitatively consistent with 
Zensus et al. (1995) and Unwin et al. (1997).

Let us next consider $\Nco{}^{\rm (SSA)}$.
This variable is roughly constant at 
$\sim 0.2 \, {\rm cm}^{-3}$ for C2-C4,
whereas it increases from $0.5 \mbox{cm}^{-3}$ at 1992.05
to $10 \mbox{cm}^{-3}$ at 1993.55 for C7.
We consider that this tendency comes from insufficient
angular resolution in particular when a component is close to the core.
We can alternatively compute $\Nco$ from
$\Nco = (K/\gamma_{\rm min}m_{\rm e}c^2)(B^2/8\pi)$,
the energy equipartition.
Reminding $K\sim 0.1$ for $\alpha \sim -0.5$,
we find that $\Nco$ computed in this way is consistent with
$\Nco{}^{\rm (SSA)}$.
 
We can compute $\Nco{}^{\rm (pair)}$, 
$\Nco{}^{\rm (kin)}$ for a pair plasma,
from equation (\ref{eq:Lkin_CPO}).
The results of $\Nco{}^{\rm (pair)}$ are presented in 
Tables \ref{tbl-2} and \ref{tbl-3},
together with the ratio 
$\Nco{}^{\rm (pair)} / \Nco{}^{\rm (SSA)}$.
It follows from Table \ref{tbl-2} that C2 and C3 are
likely dominated by a pair plasma.
It is also suggested that C4 is dominated by pair plasma unless
$L_{\rm kin}$ exceeds $10^{46.5}$ ergs/s.
Unfortunately, the errors in $B$, $\Nco{}^{\rm (SSA)}$ and
$\Nco{}^{\rm (pair)}$ cannot be calculated,
because those in $\nu_{\rm m}$ and $S_{\rm m}$ are not
presented in Zensus et al. (1995).
Furthermore, Table \ref{tbl-3} indicates that
C5 and C7 at all the four epochs are likely dominated 
by a pair plasma.
Unfortunately, the meaningful errors in 
$B$, $\Nco{}^{\rm (SSA)}$, and $\Nco{}^{\rm (pair)}$
for C5 cannot be calculated,
because its error in $\theta_{\rm d}$ (or $\xi$ in their notation) 
is not presented in Unwin et al. (1997).
Nevertheless, the results of $\Nco{}^{\rm (pair)} / \Nco{}^{\rm (SSA)}$ 
strongly suggest that the jet components of 3C 345 on parsec scales
are dominated by a pair plasma.


\section{Discussion}

In summary, we derive the proper electron number density, 
$\Nco{}^{\rm (SSA)}$, 
of a homogeneous radio-emitting component
of which spectral turnover is due to synchrotron self-absorption.
Comparing $\Nco{}^{\rm (SSA)}$ with the density derived from the 
kinetic luminosity of the jet, we can investigate whether we can
exclude the possibility of normal plasma ($e^-$-p) dominance.
Applying this method to the \lq\lq superluminal'' quasar 3C345, 
using the published spectrum data of C2, C3, C4, C5, and C7,
we find that all the five components are likely dominated 
by a pair plasma.

As demonstrated in the last part of \S 2,
the conclusion is invulnerable against the undetermined value of
$\gamma_{\rm min}$ of electrons and positrons.
However, if $\gamma_{\rm min}$ for a {\it normal} plasma
were to be significantly less than $100$,
then the possibility of a normal plasma dominance could not be
ruled out in general.
In the case of the 3C 345 jet,
equation (\ref{eq:Lkin_3C345_2}) would give 
$L_{\rm kin} \sim 10^{48} \mbox{ergs s}^{-1}$
for a normal plasma with $\gamma_{\rm min} \sim 1$.
In this case, the large kinetic luminosity  
($\sim 10^{48} \mbox{ergs s}^{-1}$) is carried by protons, because 
\begin{equation}
  \langle \gamma_- \rangle m_{\rm e}c^2
    \sim \frac{\gamma_{\rm min}}{K} m_{\rm e}c^2
    \ll  m_{\rm p}c^2
\end{equation}
holds.
Nevertheless, we consider that such a jet is unlikely,
because the protons carry about two orders of 
magnitude more energy than is seen to be dissipated as
synchrotron radiation ($\sim 10^{46} \mbox{ergs s}^{-1}$).
Electrons on parsec scales will not be cooled down so rapidly
shortly after being heated-up at the shock fronts.

It is interesting to consider the case when $\delta$ is estimated 
by other methods than the energy equipartition.
As an example, let us consider a jet motion with 
a roughly constant Lorentz factor;
Zensus et al. (1995) derived that $\Gamma \sim 10$ is
close to the smallest value that is consistent with all their
available kinematic constraints.
Such values of $\delta$ and $\varphi$ are denoted by the solid
dots in Fig. 12 of their paper and tabulated again in table~\ref{tbl-4}
in the present paper.
Using those data, we can compute $B$ and $\Nco{}^{\rm (SSA)}$ of
each component (table~\ref{tbl-4}).
For C2, we adopt $\Gamma=13$ rather than $10$,
because $\beta_{\rm app}=12.9$ for $h=1$ (or equialently $H_0=65$) gives
$\Gamma > \sqrt{1+\beta_{\rm app}^2} = 12.9$.
The results of $\Nco{}^{\rm (pair)}/\Nco{}^{\rm (SSA)}$ show again
that C2-C4 at 1982.0 are likely dominated by a pair plasma.

\acknowledgments

\appendix
\section{Derivation of the Synchrotron Self-absorption Constraints}

We assume that the parsec-scale jet close to the core
propagates conically with a half opening angle $\chi$ 
in the observer's frame. 
Then the optical depth $\tau$ for 
synchrotron self absorption is given by
\begin{equation}
  \tau_{\nu}(R) = \frac{2R\sin\chi}{\sin(\varphi+\chi)} \alpha_{\nu}, 
\label{eq:tau1}
\end{equation}
where $R$ is the distance of the position from the injection point
of the jet
and $\alpha_\nu$[1/cm] refers to the absorption coefficient.
For a small half opening angle ($\chi \ll 1$),
this equation can be approximated as
\begin{equation}
  \tau_{\nu}(R)= 2R\frac{\chi}{\sin\varphi}\alpha_{\nu}
  \label{eq:tau2}
\end{equation}
Since $\tau$ and $R \chi$ are Lorentz invariants, we obtain
\begin{equation}
  \frac{\alpha_\nu}{\sin\varphi}
  = \frac{\alpha_\nu^*}{\sin\varphi^*},
  \label{eq:LI-1}
\end{equation}
where a quantity with an asterisk is measured in the co-moving frame,
while that without an asterisk in the observer's frame.
Since $\nu \alpha_\nu$ is also Lorentz invariant, 
equation (\ref{eq:LI-1}) gives
\begin{equation}
  \frac{\sin\varphi^*}{\sin\varphi}
  = \frac{\nu}{\nu^*}
  = \frac{\delta}{1+z} .
  \label{eq:LI-3}
\end{equation}
Combining equations (\ref{eq:tau2}) and (\ref{eq:LI-3}), we obtain
\begin{equation}
  \tau_\nu
      = \frac{1+z}{\delta} \frac{2R\chi}{\sin\varphi} \alpha_\nu^* 
      = \frac{1+z}{\delta} \frac{1}{\sin\varphi}
    \frac{\theta_{\rm d} D_{\rm L}}{(1+z)^2} \alpha_\nu^* ,
  \label{eq:o-depth-2}
\end{equation}
where the angular diameter distance of the jet, $2R\chi/\theta_{\rm d}$,
is rewritten with the luminosity distance, $D_{\rm L}$, 
divided by $(1+z)^2$;
here, $\theta_{\rm d}$ is the angular diameter of the component
in the perpendicular direction of the jet propagation.
If we observe $\tau_\nu$ at the turnover frequency, $\nu_{\rm m}$,
it becomes a function of the optical thin spectral index $\alpha$,
which is tabulated in Scott and Readhead (1977).

Averaging over pitch angles of the isotropic electron power-law
distribution (eq. [\ref{eq:ele-dist}]), 
we can write down the absorption coefficient in the co-moving frame as
(Le Roux 1961, Ginzburg \& Syrovatskii 1965)
\begin{equation}
  \alpha_\nu^* = C(\alpha) r_\circ{}^2 k_{\rm e}^*
                 \frac{\nu_\circ}{\nu^*}
		 \left( \frac{\nu_{\rm B}}{\nu^*} 
                 \right)^{(-2\alpha+3)/2} ,
  \label{eq:abs-coeff}
\end{equation}
where $\nu_\circ \equiv c/r_\circ \equiv c/[e^2 / (m_{\rm e} c^2)]$
and $\nu_{\rm B} \equiv eB / (2\pi m_{\rm e}c)$.
The coefficient $C(\alpha)$ is given in Table 1 of Gould (1979).

Substituting equation (\ref{eq:abs-coeff}) into (\ref{eq:o-depth-2}),
and assuming $\alpha<0$ and $\gamma_{\rm min} \ll \gamma_{\rm max}$,
we obtain with the aid of (\ref{eq:ele-dist})
\begin{eqnarray}
  \Nco B^{-\alpha +1.5} 
  &=& \frac{m_{\rm e}c}{e^2}
      \left( \frac{e}{2\pi m_{\rm e}c} \right)^{-1.5+\alpha}
      \frac{\tau_\nu(\alpha)}{C(\alpha)}
      \frac{\gamma_{\rm min}{}^{2\alpha}}{-2\alpha}
  \nonumber \\
  & & \times \frac{(1+z)^2}{D_{\rm L}}
       \frac{\sin \varphi}{\theta_{\rm d}}
      \left( \frac{1+z}{\delta} \right)^{-\alpha+1.5}
      \nu^{-\alpha+2.5}.
  \label{eq:lineB-1}
\end{eqnarray}
Evaluating $\nu$ at the turnover frequency, $\nu=\nu_{\rm m}$,
and combining with equation (\ref{eq:lineA}),
we obtain $\Nco$ presented in equation (\ref{eq:Nmin}),
which equals $(\gamma_{\rm min}m_{\rm e}c^2){}^{2\alpha}/(-2\alpha)$
times $N_0$ given in equation~(3) in Marscher (1983).
It is noteworthy that electron number density in the observer's frame
can be obtained if we multiply $(1+z)/\delta$ on $\Nco$.


\section{Kinetic luminosity inferred from core-position offset}

In this appendix, we deduce the kinetic luminosity of a jet
from its core-position offset due to synchrotron self-absorption.
This method was originally developed by Lobanov (1988).
However, our results somewhat differs from his results;
therefore, we explicitly describe the derivation 
so that the readers can check it.

\subsection{Scaling Law}

First, we assume that $\Nco$ and $B$ scale on $r$ 
in the following manner:
\begin{eqnarray}
  N_e^{\ast}= N_1 r^{-n}, \quad   B = B_1 r^{-m}, 
  \label{eq:scaling}
\end{eqnarray}
where $N_1$ and $B_1$ refer to the values of
$\Nco$ and $B$ at $r_1 = 1$ pc, respectively;
$r \equiv R/r_1$.
Introducing dimensionless variables
\begin{eqnarray}
  x_{\rm N} \equiv r_1 r_\circ{}^2 N_1
  \nonumber \\
  x_{\rm B} \equiv \nu_{\rm B_1}/\nu_\circ
    = \frac{eB_1}{2\pi m_{\rm e}c},
  \label{eq:def_x}
\end{eqnarray}
and utilizing equation (\ref{eq:abs-coeff}), 
we obtain from the left equality in equation (\ref{eq:o-depth-2})
\begin{equation}
  \tau_\nu = C(\alpha) \frac{2\chi}{\sin\varphi}
             \frac{-2\alpha}{\gamma_{\rm min}{}^{2\alpha}}
             \left(\frac{1+z}{\delta}\right)^{-\epsilon}
             \left(\frac{\nu}{\nu_\circ}\right)^{-1-\epsilon}
             r^{1-n-m\epsilon}
             x_{\rm N} x_{\rm B}{}^\epsilon,
  \label{eq:tau4}
\end{equation}
where $\epsilon \equiv 3/2 -\alpha$.

At a given frequency $\nu$, 
the flux density will peak at the position where $\tau_\nu$ becomes unity.
Thus setting $\tau=1$ and solving equation (\ref{eq:tau4}) 
for $r$, we obtain the distance from the VLBI core 
observed at frequency $\nu$ from the central engine as 
\begin{equation}
  r(\nu) = \left( x_{\rm B}{}^{k_{\rm b}} F \frac{\nu_\circ}{\nu}
           \right)^{1/k_{\rm r}}
  \label{eq:core_rad} 
\end{equation}
where 
\begin{equation}
  F(\alpha) \equiv \left[ C(\alpha) \frac{2\chi}{\sin\varphi}
                          \frac{-2\alpha}{\gamma_{\rm min}{}^{2\alpha}}
                          \left(\frac{\delta}{1+z} \right)^{\epsilon} 
                          x_{\rm N} 
                   \right]^{1/(\epsilon+1)}
\end{equation}
\begin{equation}
  k_{\rm b} \equiv \frac{3-2\alpha}{5-2\alpha},
\end{equation}
\begin{equation}
  k_{\rm r} \equiv \frac{(3-2\alpha)m+2n-2}{5-2\alpha}. 
\end{equation}

\subsection{Core-Position Offset}

If we mesure $r(\nu)$ at  two different frequencies
(say $\nu_{\rm a}$ and $\nu_{\rm b}$),
equation (\ref{eq:core_rad}) gives the dimensionless, projected 
distance of
$r(\nu_{\rm a})-r(\nu_{\rm b})$ as 
\begin{equation}
  \Delta r_{\rm proj} 
  = \left[ r(\nu_{\rm a}) - r(\nu_{\rm b}) \right] \sin\varphi
  = (x_{\rm B}{}^{k_{\rm b}} F \nu_\circ)^{1/k_{\rm r}}
    \frac{\nu_{\rm b}^{1/k_{\rm r}} - \nu_{\rm a}^{1/k_{\rm r}}}
    {\nu_{\rm a}^{1/k_{\rm r}} \nu_{\rm b}^{1/k_{\rm r}}}
    \sin\varphi.
  \label{eq:del_jet}
\end{equation}
Defining the core-position offset as
\begin{equation}
  \Omega_{r \nu} \equiv 
    r_1 \Delta r_{\rm proj}
    \frac{\nu_{\rm a}^{1/k_{\rm r}} \nu_{\rm b}^{1/k_{\rm r}}}
         {\nu_{\rm b}^{1/k_{\rm r}} - \nu_{\rm a}^{1/k_{\rm r}}},
  \label{eq:def_CPO}
\end{equation}
we obtain
\begin{equation}
  \frac{\Omega_{r\nu}}{r_1}
  = (x_{\rm B}^{k_{\rm b}} F \nu_\circ )^{1/k_{\rm r}} \sin\varphi
  \label{eq:CPO}
\end{equation}

To express $x_{\rm B}$ in terms of $x_{\rm N}$ and 
$\Omega_{r\nu}$, we can invert equation (\ref{eq:CPO}) as
\begin{equation}
  x_{\rm B} = \left( \frac{\Omega_{r\nu}}{r_1 \sin\varphi}
              \right)^{k_{\rm r}/k_{\rm b}}
              (F \nu_\circ)^{-1/k_{\rm b}}.
  \label{eq:xB}
\end{equation}
Note that $x_{\rm N}$ is included in $F=F(\alpha)$.

Setting $\nu_{\rm b} \rightarrow \infty$ in equation (\ref{eq:del_jet}), 
we obtain the absolute distance of the VLBI core
measured at $\nu$ from the central engine as
\begin{equation}
  r_{\rm core} (\nu) = \frac{\Omega_{r \nu}}{r_1 \sin \varphi}
                       \nu^{-1/k_{\rm r}}.
\end{equation}
That is, once $\Omega_{r\nu}$ is obtained from multi-frequency VLBI
observations,
we can deduce the distance of the synchrotron-self-absorbing VLBI core 
from the central engine,
assuming the scaling laws of $\Nco$ and $B$ as equation (\ref{eq:scaling}).

We next represent $x_{\rm N}$ and $x_{\rm B}$ 
(or equivalently, $N_1$ and $B_1$) as a function of $\Omega_{r\nu}$.
To this end, we relate $\Nco$ and $B$ as follows:
\begin{equation}
  \Nco \gamma_{\min} m_{\rm e} c^2 = K \frac{B^2}{8 \pi}.
  \label{eq:equiP_0}
\end{equation}
When an energy equipartition between the radiating particles
and the magnetic field holds,
equation (\ref{eq:ele-dist}) gives for $\alpha=-0.5$
\begin{equation}
  K = \frac{1}{\ln(\gamma_{\rm max}/\gamma_{\rm min})}
    \sim 0.1,
  \label{eq:K_1}
\end{equation}
whereas for $\alpha<-0.5$
\begin{equation}
  K = \frac{2\alpha+1}{2\alpha}
      \frac{ \gamma_{\rm max}{}^{2\alpha}
            -\gamma_{\rm min}{}^{2\alpha}}
           { \gamma_{\rm max}{}^{2\alpha+1}
            -\gamma_{\rm min}{}^{2\alpha+1}}.
  \label{eq:K_2}
\end{equation}

Substituting $N_{\rm e}^* = N_1 r^{-2}$ and $B = B_1 r^{-1}$ into 
(\ref{eq:equiP_0}),
and replacing $N_1$ and $B_1$ with $x_{\rm N}$ and $x_{\rm B}$,
we obtain
\begin{equation}
  x_{\rm N}= \frac{\pi}{2} \frac{K}{\gamma_{\rm min}}
             \frac{r_1}{r_\circ}
             x_{\rm B}{}^2
  \label{eq:equiP_1}
\end{equation}
It is noteworthy that the assumptions of $n=2$ and $m=1$, 
which results in $k_{\rm r}=1$,
guarantees the energy equipartition at an arbitrary distance, $r$.

Combining equations (\ref{eq:xB}) and (\ref{eq:equiP_1}),
we obtain
\begin{eqnarray}
  x_{\rm B}
  &=& \left( \frac{\Omega_{r\nu}/\nu_\circ}{r_1 \sin\varphi}
      \right)^{(5-2\alpha)/(7-2\alpha)}
  \nonumber \\
  && \hspace{-0.5 truecm} \times 
     \left[ \pi C(\alpha) \frac{\chi}{\sin\varphi}
             \frac{K}{\gamma_{\rm min}}
             \frac{r_1}{r_\circ}
             \frac{-2\alpha}{\gamma_{\rm min}{}^{2\alpha}}
             \left(\frac{\delta}{1+z}\right)^\epsilon
      \right]^{-2/(7-2\alpha)}.
  \label{eq:sol_xB}
\end{eqnarray}
The particle number density, $x_{\rm N}$, can be
readily computed from equation (\ref{eq:equiP_1}).

\subsection{Kinetic luminosity}

We can now relate the kinetic luminosity 
with the core-position offset.
The factor $N_{\rm e}{}^* R^2$ in equation (\ref{eq:Lkin})
can be expressed in terms of $x_{\rm N}$ and hence $x_{\rm B}$ as
\begin{eqnarray}
  \Nco R^2 
    &=& N_1 r_1{}^2 = \frac{r_1}{r_\circ{}^2} x_{\rm N}
  \nonumber \\
    &=& \frac{\pi}{2} \frac{K}{\gamma_{\rm min}}
             \frac{r_1{}^2}{r_\circ{}^3} x_{\rm B}{}^2.
\end{eqnarray}

For a pure pair plasma, 
we obtain $\langle\gamma_+\rangle= \langle\gamma_-\rangle$ 
and $m_+ = m_{\rm e}$.
Therefore, for a conical geometry, we can put $R_\perp=R\chi$
in equation (\ref{eq:Lkin}) to obtain equation (\ref{eq:Lkin_CPO}),
where 
$C_{\rm kin}= \pi^2 \langle\gamma_-\rangle m_{\rm e}c^3/\gamma_{\rm min}$.

In the same manner, 
for a normal plasma, 
we have $\langle\gamma_+\rangle=1$ and $m_+ = m_{\rm p}$. 
In this case, we obtain 
$C_{\rm kin}= \pi^2 m_{\rm p}c^3/(2\gamma_{\rm min})$.

\clearpage
 
\begin{deluxetable}{cccc}
\footnotesize
\tablecaption{Magnetic field and electron density 
		of each component \label{tbl-2}}
\tablewidth{0pt}
\tablehead{
\colhead{component}
	& \colhead{C2}		& \colhead{C3}
	& \colhead{C4}
	\nl
epoch	& 1982.0		& 1982.0
	& 1982.0
} 
\startdata
$\rho h$ [mas] \tablenotemark{a}	
	& 4.9/0.65	& 2.2/0.65	& 0.40/0.65	\nl
$\beta_{\rm app}h$\tablenotemark{a}
	& 8.4/0.65	& 6.0/0.65	& 4.0/0.65	\nl
$\nu_{\rm m}$ [GHz]\tablenotemark{a}
	& 1.5		& 2.6		& 14.6		\nl
$S_{\rm m}$ [Jy]\tablenotemark{a}
	& 2.0		& 2.1		& 7.6		\nl
$\alpha$\tablenotemark{a}
	& $-$0.6	& $-$0.7	& $-$0.3	\nl
$\theta_{\rm d}$ [mas]\tablenotemark{a}
	& 2.15		& 0.97		& 0.29		\nl
$\delta_{\rm min}$ \tablenotemark{a}
	& 2.1		& 3.6		& 14.3		\nl
\hline
$\delta_{\rm eq}$  \tablenotemark{b}
	& 6.7		& 13		& 17		\nl
$\Gamma_{\rm eq}$\tablenotemark{b}
	& 16		& 9.8		& 9.6		\nl
$\varphi_{\rm eq}$ [rad]\tablenotemark{b}
	& 0.12		& 0.072		& 0.039		\nl
$B$ [mG]  \tablenotemark{b}
	& 5.7		& 6.9		& 18		\nl
$\Nco{}^{\rm (SSA)}$  [cm${}^{-3}$] 
 \tablenotemark{b}
	& 0.11		& 0.33		& 0.19		\nl
$\Nco{}^{\rm (pair)}$  [cm${}^{-3}$] \tablenotemark{b}
	& $0.095 L_{46.5}$
	& $1.3 L_{46.5}$
	& $16 L_{46.5}$	
	\nl
$\Nco{}^{\rm (pair)}/\Nco{}^{\rm (SSA)}$  \tablenotemark{b}
	& $0.86  L_{46.5}$
	& $4.0  L_{46.5}$
	& $80	L_{46.5}$	\nl
\hline
$e^\pm$ dominated?
	& {\bf likely yes}
	& {\bf likely yes}
	& {\bf maybe  yes}
\enddata
\tablenotetext{a}{From Zensus et al. (1995).}
\tablenotetext{b}{The values for $h=1.0$ are presented.
 Kinetic luminosity is normalized as 
 $L_{46.5} \equiv L_{\rm kin}/10^{46.5}{\rm ergs}\cdot{\rm s}^{-1}$
 (see text).      }
 
\end{deluxetable}

\begin{deluxetable}{cccccc}  
\footnotesize
\tablecaption{Mangnetic field and electron density of each 
		component \label{tbl-3}}
\tablewidth{0pt}
\tablehead{
\colhead{component}
	& \colhead{C5}
	& \colhead{C7}		& \colhead{C7}
	& \colhead{C7}		& \colhead{C7}
	\nl
epoch	& 1990.55
	& 1992.05		& 1992.67
	& 1993.19		& 1993.55
} 
\startdata
$\rho h$ [mas] \tablenotemark{b}	
	& 1.75/0.65	& 0.14/0.65	& 0.22/0.65
	& 0.38/0.65	& 0.52/0.65	\nl
$\beta_{\rm app}h$
	& 5.7/0.65 \tablenotemark{a}
	& 1.8/0.65 \tablenotemark{b}	
	& 3.9/0.65 \tablenotemark{b}
	& 6.8/0.65 \tablenotemark{b}	
	& 9.4/0.65 \tablenotemark{b}
	\nl
$\nu_{\rm m}$ [GHz]\tablenotemark{b}
	& $2.7  \pm 0.5$
	& $12.8 \pm 0.5$
	& $12.5 \pm 1.0$
	& $11.6 \pm 0.5$
	& $11.0 \pm 1.5$ \nl
$S_{\rm m}$ [Jy]\tablenotemark{b}
	& $3.2 \pm 0.5$
	& $4.6 \pm 0.5$
	& $7.0 \pm 0.5$
	& $5.1 \pm 0.5$
	& $3.1 \pm 0.5$ \nl
$\alpha$\tablenotemark{b}
	& $-$0.75	& $-$0.75	& $-$0.75	& $-$0.75
	& $-$0.75	\nl
$\theta_{\rm d}$ [mas]\tablenotemark{b}
	& 0.80
	& $0.20 \pm .04$	& $0.35 \pm .02$
	& $0.41 \pm .02$	& $0.38 \pm .02$ \nl
$\delta_{\rm min}$ \tablenotemark{b}
	& $8.0  \pm 3.5$	
	& $11.7	\pm 4.1$
	& $6.5	\pm 0.9$
	& $5.5	\pm 0.6$
	& $4.0	\pm 1.1$	\nl
\hline
\rule{0.0truecm}{0.90 truecm}
$\delta_{\rm eq}$ \tablenotemark{c}
	& 33	
	& $39	\displaystyle{+43  \atop -24 }$
	& $15	\displaystyle{+6   \atop -5  }$
	& $8.4	\displaystyle{+2.5 \atop -2.3}$
	& $7.1	\displaystyle{+6.7 \atop -4.7}$
	\nl
\rule{0.0truecm}{0.90 truecm}
$\Gamma_{\rm eq}$\tablenotemark{c}
	& 18
	& $20	\displaystyle{+21  \atop -12 }$
	& $8.8	\displaystyle{+2.6 \atop -2.1}$
	& $11.0	\displaystyle{+0.8 \atop -0.8}$
	& $25	\displaystyle{+19  \atop -10}$
	\nl
\rule{0.0truecm}{0.90 truecm}
$\varphi_{\rm eq}$ [rad]\tablenotemark{c}
	& $.022$
	& $.008	\displaystyle{+.014 \atop -.007}$
	& $.05	\displaystyle{+.04  \atop -.03}$
	& $.12	\displaystyle{+.02 \atop -.03}$
	& $.11	\displaystyle{+.02 \atop -.04}$
	\nl
\rule{0.0truecm}{0.90 truecm}
$B$ [mG]  \tablenotemark{c}
	& $4.4$	
	& $19	\displaystyle{+11 \atop -11}$	
	& $31	\displaystyle{+12 \atop -12}$	
	& $43	\displaystyle{+12 \atop -12}$	
	& $62	\displaystyle{+61 \atop -39}$	
	\nl
\rule{0.0truecm}{0.90 truecm}
$ \Nco{}^{\rm (SSA)}$ [cm${}^{-3}$]\tablenotemark{c}
	& $0.09$
	& $0.5	\displaystyle{+1.9 \atop -0.4}$
	& $4.5	\displaystyle{+6.6 \atop -2.9}$
	& $11	\displaystyle{+5   \atop -4}$
	& $12	\displaystyle{+10   \atop -8}$
	\nl
\rule{0.0truecm}{0.90 truecm}
$ \Nco{}^{\rm (pair)} $ [cm${}^{-3}$]\tablenotemark{c}
	& $0.63				      L_{46.5}$
	& $11	\displaystyle{+17 \atop -8}  L_{46.5}$
	& $14	\displaystyle{+7  \atop -6} L_{46.5}$
	& $5.9	\displaystyle{+1.4 \atop -1.8}  L_{46.5}$
	& $1.9	\displaystyle{+1.9 \atop -1.6}   L_{46.5}$
	\nl
\rule{0.0truecm}{0.90 truecm}
$ \Nco{}^{\rm (pair)} / \Nco{}^{\rm (SSA)} $  \tablenotemark{c}
	& $7.2					 L_{46.5}$
	& $23   \displaystyle{+40  \atop -12 }	 L_{46.5}$
	& $3.0  \displaystyle{+1.9 \atop -1.2}	 L_{46.5}$
	& $0.53 \displaystyle{+0.44 \atop -0.27} L_{46.5}$
	& $0.17 \displaystyle{+0.81 \atop -0.15} L_{46.5}$
	\nl
\hline
  $e^\pm$ dominated?
	& {\bf likely yes}
	& {\bf likely yes}
	& {\bf likely yes}
	& {\bf likely yes}
	& {\bf likely yes} \nl
\enddata
\tablenotetext{a}{From Unwin \& Wehrle (1992).}
\tablenotetext{b}{From Unwin et al. (1997).
  Errors are nominally 1 $\sigma$ but are dominated by 
  systematic errors, which are included in the estimate.}
\tablenotetext{c}{The values for $h=1$ are presented.
  $L_{46.5} \equiv L_{\rm kin}/10^{46.5}{\rm ergs}\cdot{\rm s}^{-1}$.
  Errors are $90\%$ confidence regions for a single parameter of 
  interest.}
\end{deluxetable}

\begin{deluxetable}{cccc}
\footnotesize
\tablecaption{Electron densities when $\Gamma$ is given \label{tbl-4}}
\tablewidth{0pt}
\tablehead{
\colhead{component}
	& \colhead{C2}		& \colhead{C3}
	& \colhead{C4}
	\nl
epoch	& 1982.0		& 1982.0
	& 1982.0
} 
\startdata
$\rho$ [mas]
	& 4.9/0.65	& 2.2/0.65	& 0.40/0.65	\nl
$\beta_{\rm app} h$
	& 8.4/0.65	& 6.0/0.65	& 4.0/0.65	\nl
$\Gamma$
	& 13		& 10		& 10		\nl
\hline
$\delta$
	& 14		& 13		& 18		\nl
$B$ [mG]  \tablenotemark{a}
	& 11		& 7.2		& 19		\nl
$\Nco{}^{\rm (SSA)}$  [cm${}^{-3}$] \tablenotemark{a}
	& 0.0055	& 0.77		& 1.2		\nl
$\Nco{}^{\rm (pair)}$  [cm${}^{-3}$] \tablenotemark{a}
	& $0.15 L_{46.5}$
	& $1.3	L_{46.5}$	
	& $14	L_{46.5}$
	\nl
$\Nco{}^{\rm (pair)} / \Nco{}^{\rm (SSA)}$  \tablenotemark{a}
	& $27	L_{46.5}$
	& $1.6	L_{46.5}$
	& $11	L_{46.5}$
	\nl
\hline
$e^\pm$ dominated?
	& {\bf likely yes}
	& {\bf likely yes}
	& {\bf likely yes}
\enddata
\tablenotetext{a}{$h=1$ is assumed.
 $L_{46.5} \equiv L_{\rm kin}/10^{46.5}{\rm ergs}\cdot{\rm s}^{-1}$. }
 
\end{deluxetable}

%
%

\clearpage


\end{document}